# Domain-specific Knowledge Graphs: A survey

Bilal Abu-Salih

King Abdullah II School of Information Technology

The University of Jordan

b.abusalih@ju.edu.jo

## Abstract

Knowledge Graphs (KGs) have made a qualitative leap and effected a real revolution in knowledge representation. This is leveraged by the underlying structure of the KG which underpins a better comprehension, reasoning and interpretation of knowledge for both human and machine. Therefore, KGs continue to be used as the main means of tackling a plethora of real-life problems in various domains. However, there is no consensus in regard to a plausible and inclusive definition of a domain-specific KG. Further, in conjunction with several limitations and deficiencies, various domain-specific KG construction approaches are far from perfect. This survey is the first to offer a comprehensive definition of a domain-specific KG. Also, the paper presents a thorough review of the state-of-the-art approaches drawn from academic works relevant to seven domains of knowledge. An examination of current approaches reveals a range of limitations and deficiencies. At the same time, uncharted territories on the research map are highlighted to tackle extant issues in the literature and point to directions for future research.

## 1. Introduction

KGs, one of the key trends which are driving the next wave of technologies [1], have now become a new form of knowledge representation, and the cornerstone of several applications ranging from generic to specific industrial usage cases [2]. The ever-increasing interest in this technology is due to its underlying abstract structure which effectively facilitates domain conceptualization and data management, and its usage as the main driver of several Artificial Intelligence applications. In particular, the KG depicts an integrated collection of real-world entities which are connected by semantically-interrelated relations. In this respect, data are given formal semantics via data annotation and manipulation in a machine-readable format, thereby reducing ambiguity and deriving meaningful information that is specific to an application's domain. Therefore, the incorporation of KGs has extended the existing data models depicted by domain ontologies and established a new form of data analytics that is able to capture semantically-interconnected, large-scale data sets.

Beyond the generic and open-world KGs such as Google KG [3], most of the current KGs are domain-specific, with certain underlying ontologies in their design [4]. Because of the lack of a 'one-size fits all' schema or ontology that can be applied to address real-life problems, efforts to establish, polish, and augment domain-specific KGs are continuing to be made in several domains of knowledge [5]. However, this ongoing interest in domain-specific KGs raises questions about their quality and robustness, and whether adequate evaluation measures have been applied, particularly to those KGs derived from data sources of inconsistent quality. Also, the dynamic nature of domain knowledge is highly correlated to contextual situations, and various facts that describe entities might change over time. Neglecting the dynamic nature of knowledge diminishes the quality and correctness of facts represented by KGs, and could lead to poor decision making that is based merely on such data sources. Therefore, it is important that a comprehensive review be conducted of the current state-of-the-art approaches for domain-specific KG construction so as to highlight such issues and address them with viable solutions.

In this survey, we provide an inclusive definition to domain-specific KGs. Further, we discuss various notable KG construction approaches in seven domains of knowledge. These approaches are reviewed and a summary is provided for each domain showing how the KG in each case has been constructed, the resources used to construct the KG, whether any of the KG embedding techniques have been incorporated, the measures used to evaluate the KG construction approach, and the limitations and shortcomings of each approach. This survey paper is different from other similar researches that tend to either focus on generic and domain-independent KGs such as [6-8], or only briefly discuss domain-specific KGs [5, 9]. To the best of our knowledge, this is the first attempt to provide both an inclusive definition of a domain-specific KG as well as a thorough analysis of various domain-based KG construction approaches. Further, this paper provides a summary of the main issues derived from the conducted analysis. At the same time, several improvements, recommendations, and opportunities are suggested to address these limitations which point to directions for future research. The contributions of this paper are:

- To the best of our knowledge, this is the first paper to provide an inclusive definition of a domain-specific KG and the first comprehensive survey of domain-specific KGs.
- We conduct a thorough analysis of more than 140 papers on KG construction approaches, covering seven domains.
- The paper highlights research gaps in the area of domain-specific KG construction and suggests venues for future research.

The rest of this paper is organised as follows: Section 2 discusses the methodology adopted in this study. Section 3 establishes the necessary ground for this work by including important preliminaries and relevant terminologies. Section 4 analyses the KG construction approaches in seven domains and presents an in-depth discussion of the findings as well as the identified research gaps. Section 6 concludes the paper.

## 2. Methodology

As the first step, seven domains were chosen and the articles for review were collected accordingly. The domains are: healthcare, education, ICT, science and engineering, finance, society and politics, and travel. Articles relevant to any of these domains were obtained by searching recent volumes of both conference proceedings of relevant series (such as ACM SIGKDD, ACM WSC, WWW, ICWE, ISWC, etc.) and high-quality journals (such as Knowledge-based Systems, Expert Systems with Applications, IEEE Access, Journal of Web Semantic, etc.). Furthermore, we used keywords such as "knowledge graph for engineering", "knowledge graph for healthcare", etc. to search for articles in Google Scholar. Although the articles chosen for each domain intuitively depict the KG construction approaches to the domain of interest, it was noted that some papers were concerned with more than one high-level domain of knowledge. Those papers dealing with interrelated domains are positioned under the domain umbrella that is explicitly indicated by the papers' authors.

We examined more than 140 research articles that appeared between 2016 to 2020 in high-quality computer science and information systems publication venues. Figure 1 indicates the number of papers collected for the review. As seen in the figure, the interest in domain-specific KGs has increased dramatically over recent years.

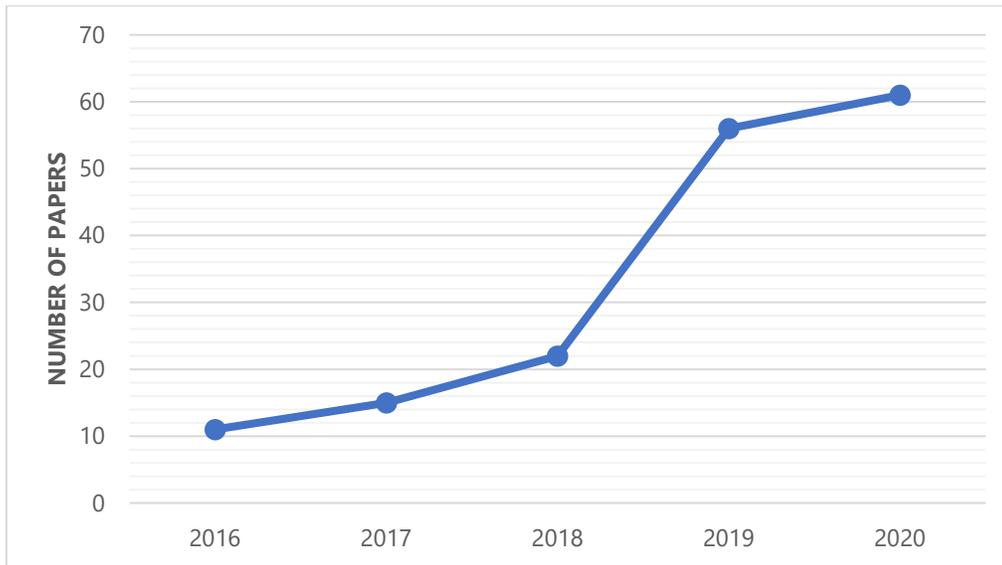

Figure 1. Number of studied papers per year.

Our study is different from other similar works. For example, the current important researches in this arena, such as [6-8], either focus on generic and domain-independent KGs, or only touch upon and do not discuss in depth domain-specific KGs [5, 9]. To the best of our knowledge, this is the first attempt to provide both an inclusive definition of the term 'domain-specific KG' as well as a thorough analysis of various domain-based KG construction approaches. Furthermore, the review reveals the shortcomings of current approaches and proposes solutions to address them.

## 3. Preliminaries

### 3.1 Generic Knowledge Graphs

Generic KGs (a.k.a. open-world, cross-domain, or domain-independent) were constructed long before the term "knowledge graph" was coined, and such constructions have been ongoing. In fact, since the invention of the Semantic Web, generic KGs have been associated with linked data, being natural representations of the interlinking of entities [10]. Nevertheless, the term has gained much momentum recently as it has given rise to new computing paradigms by shifting from traditional databases to knowledge databases [11]. Ironically, there is no consensus on the definition of the term despite the few attempts to provide a reasonable description. For example, Ehrlinger and Wöß [10] perceive the KG as the process of acquiring and correlating knowledge to an ontology and applying a reasoner to infer knowledge. A further technical depiction of the term is provided by Wang et al. [8] who conceive the KG as a multidimensional graph comprising entities/nodes and relations/edges.

A KG is commonly described as a directed graph ($G$), where $G = (V, E)$. This notation depicts the relationship between vertices ($V$) of the graph and edges ($E$) between these vertices. The vertices represent the set of real-world entities and the edges represent the relationships between these entities. Vertices/entities/nodes are interconnected using relations which are the edges of the graph, and facts are commonly represented as an RDF[1] triple (*subjects, predicate, object*) or (*head, relation, tail*), and denoted as $< h, r, t >$. Intuitively, two entities connected by a relation form a fact in the KG. For example, Figure 2 gives an example of a KG representation in terms of entities and relations.

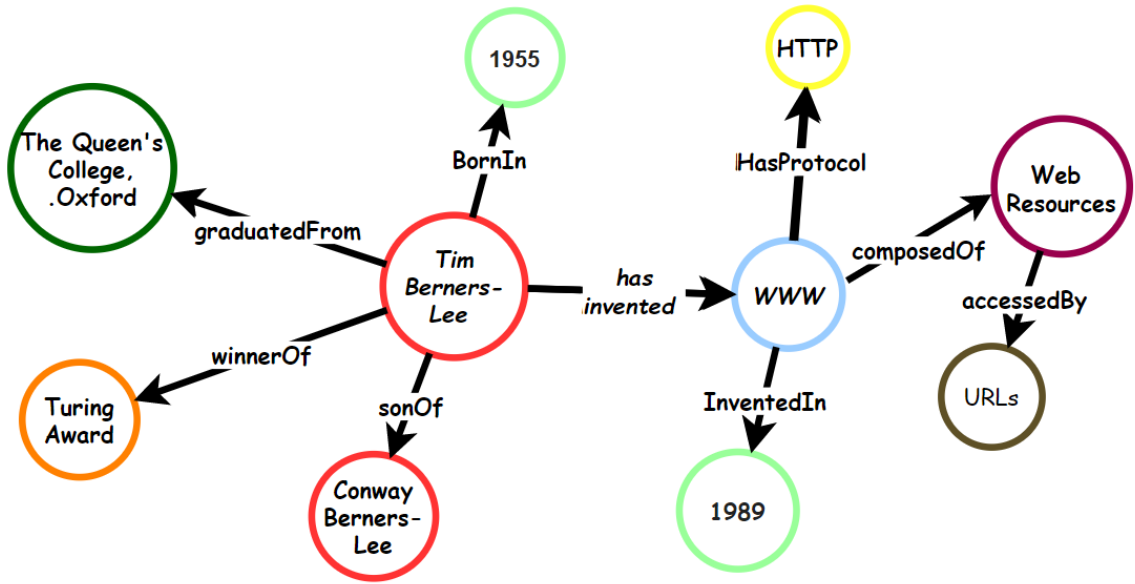

*Figure 2: An example of entities and relations in a KG*

Several facts (triples) can be inferred from this KG. For example, the facts "*Tim Berners-Lee has invented WWW*" comprises two entities/nodes, namely "*Tim Berners-Lee*" and "*WWW*", and the relation "*has invented*" forms the triple "*Tim Berners-Lee, hasInvented, WWW*".

Examples of various steadily evolving open-world KGs include: BabelNet[2], YAGO[3], Cyc[4], NELL[5], CliGraph[6] and DBPedia[7] knowledge base. In fact, several of these massive publicly-available data islands have been harvested from the Web as key sources of knowledge to benefit numerous Artificial Intelligence and smart systems [12], such as recommender systems [13], decision support systems (DSSs) [14], and intelligent QA systems [15]. Table 1 presents statistics for popular generic KGs. The statistics include #instances, #facts, #types (indicating the number of classes in the designated schema), and #relations.

---

[1] https://www.w3.org/TR/rdf11-concepts/
[2] https://babelnet.org/
[3] http://www.foaf-project.org/
[4] https://www.cyc.com/
[5] http://rtw.ml.cmu.edu/rtw/kbbrowser/
[6] http://caligraph.org/ontology/Scientist
[7] https://wiki.dbpedia.org/

*Table 1: Statistics for some of the open world KGs [16]*

|          | Instances/entities | Facts       | Types/classes | Relations |
|----------|-------------------:|------------:|--------------:|----------:|
| **DBpedia**  | 5,044,223 | 400         | 760           | 1355      |
| **YAGO**     | 6,349,359 | 479,392,870 | 819,292       | 77        |
| **Wikidata** | 52,252,549| 732,420,508 | 2,356,259     | 6,236     |
| **BabelNet** | 7,735,436 | 178,982,397 | 6,044,564     | 22        |
| **Cyc**      | 122,441   | 2,229,266   | 116,821       | 148       |
| **NELL**     | 5,120,688 | 60,594,443  | 1,187         | 440       |
| **CaLiGraph**| 7,315,918 | 517,099,124 | 755,963       | 271       |
| **Voldemort**| 55,861    | 693,428     | 621           | 294       |

### 3.2 Domain-specific Knowledge Graphs

Despite the extensive use of the generic and open-world KGs to tackle a wide variety of domain-independent tasks, constructing KGs from domain corpora to address domain-specific problems is greatly important [17]. This is because domain-specific KGs have relevant and semantically interlinked applications with domain-specific problems. Moreover, domain-specific KGs also lack a well-established, consensual and comprehensive definition, which is not surprising given that this is still a comparatively new territory and an under-explored frontier [18]. Nevertheless, some studies perceive the domain-specific KG as a special type of KG that is used to represent a specific and complex domain [4, 19, 20]. Others reported that domain-specific KGs are the result of the process of enriching an underlying domain ontology [5]. This inability to provide an inclusive definition to the domain-specific KG has driven us to frame the following definition of this term:

> "*Domain Knowledge Graph is an explicit conceptualisation to a high-level subject-matter domain and its specific subdomains represented in terms of semantically interrelated entities and relations*".

This comprehensive definition addresses three core aspects, namely: (i) formal conceptualisation: which indicates the logical design of the KG depicted by a specific and predefined domain ontology established to capture the domain of interest either in its generic (high-level) sense or in its specific subdomains; (ii) subject-matter domain: this frames the domain-specific KG to be firmly contextualised to address a particular subject-matter knowledge; and (iii) semantically interrelated entities and relations, which indicate the physical design of the domain-specific KG depicted as a labelled graph in which the semantics of data is enriched with a specific conceptual representation of entities and relationships between these entities.

### 3.3 Knowledge Graph Construction

KG was introduced as an efficient and smart approach to tackling the continuous propagation of various forms of unstructured text (e.g. Web data) and other structured or semi-structured sources [6]. The construction of a KG can be perceived as a paradigm which comprises different perspectives. We have examined the literature on KG construction approaches and we design a taxonomy to highlight the key aspects of KG construction. Figure 3 illustrates the taxonomy for KG construction which is mainly categorised based on the level of knowledge extraction, the type of knowledge base, and the incorporated construction method.

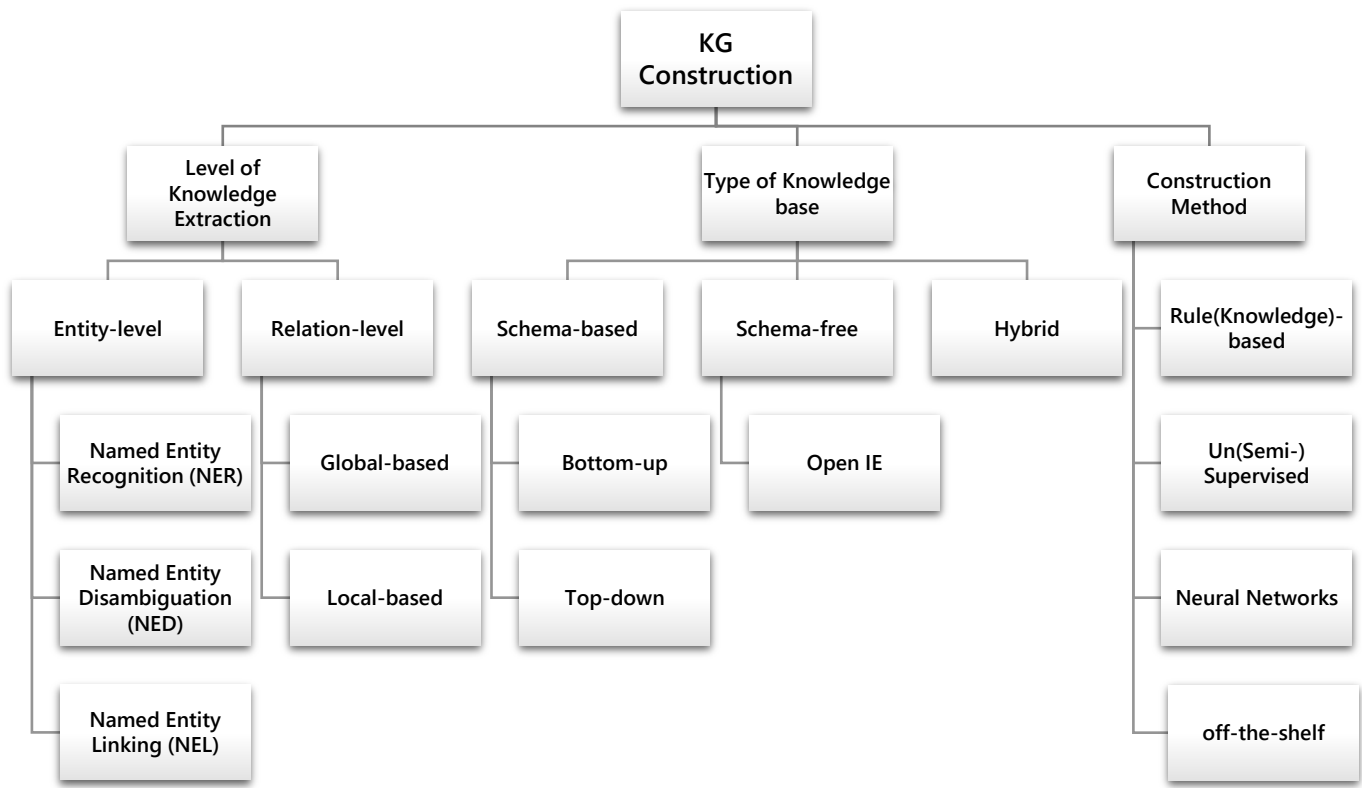

*Figure 3: A taxonomy for KG construction*

In terms of knowledge extraction, KG construction is the process of extracting entities and relationships between these entities. Entity extraction comprises three key tasks [21], namely: i) Named Entity Recognition (NER) which involves the process of finding individuals, organisation, locations, events, and other entities from (un )(semi-)structured data sources; ii) Named Entity Disambiguation (NED) which aims to eliminate the ambiguity of an inferred entity by mapping it to the factual real-world entity; and iii) Named Entity Linking (NEL) which assigns a unique IRI identified to the disambiguated entity. The purpose of relation extraction is to discover the semantic relationships between the identified and disambiguated entities. Relation extraction can be done by either local-based or global-based relation extraction methods. The former indicates the mention-level relation which is commonly inferred form short textual contents, and the latter aims to infer those relations that span several local-based relations. The reader can obtain further details on these tasks in relevant literature such as [22, 23].

Another perspective conceives that the KG construction depends on whether a predefined ontology schema is used (schema-based), or the KG is constructed with no predefined schema (schema-free) [24], or if the construction comprises a hybrid of schema-based and schema-free approaches. The first methodology (schema-based) can be categorised according to two groups based on the selection of data sources and ontology [25, 26] namely: (i) the bottom-up method whereby the structural nature of an ontology is incorporated to build the KG (e.g. Wikipedia is established by using the predefined ontology model, i.e. DBpedia [27]); and (ii) the top-down method where the ontology schema is inferred from the underlying structured data (e.g. YAGO) [28]) or build taxonomies (hierarchy) are constructed based on information from the Web [29]. Schema-free approaches are open information extraction techniques which rely on the openness of the Web; hence, the information is collected using various knowledge extraction techniques with no consideration given to creating a unified ontology design (e.g. OpenIE [30]). Hybrid knowledge-based approaches: are the techniques whereby the knowledge is obtained based on a predetermined ontology, as well as facts gathered from the Web (e.g. KnowledgeVault [25], NELL [31]).

The third thread of research sees the KG construction as being based on the technical solutions and methods used to mine the Web and other data silos and repositories to deduce entities and relations, thereby constructing the KG. Various approaches were proposed in the literature to address this issue. These fall under four key categories [6, 21, 23]: (1) Knowledge-based approaches: these methods mainly involve domain experts to establish human-crafted rules (Association Rules) by incorporating domain-knowledge and lexicon resources. (2) Learning-based methods: these methods encompass (un)(semi-)supervised techniques. Examples of these methods that are employed for entities recognition and relations extraction are Conditional Random Field (CRF), Hidden Markov Models (HMM), Support Victor Machine (SVM), Naïve Bayes (NB), Logistic Regression (LR), and Decision Trees, and bootstrapped methods. (3) Neural network models: these methods incorporate neural networks and deep learning to infer features. These methods require neither domain ontologies nor domain-specific corpora, thereby making them suitable for domain-independent KGs. Examples of these methods Convolutional Neural Network (CNN), Recurrent Neural Network (RNN), Bidirectional Long Short-Term Memory (BiLSTM). (4) Off-the-shelf NLP tools: those are commonly commercial/open source ready-to-use tools which might embed one or more of the aforementioned approaches to analyse textual content and extract entities and relations. Examples of these tools are spaCy[8], Stanford CoreNLP[9], AllenNLP [10], and IBM Watson NLU[11].

## 3.4 Knowledge Graph Embedding

KG Embedding (KGE) is the process of creating propositional feature vector representations of the constituents of a KG (entities and relationships) [8] so as to apply numeric techniques that produce scalable and effective results [32]. KGE techniques can simplify the resolution of various complex real-life graph problems for which conventional graph presentation (i.e., adjacency matrix) is inadequate and inferior. Hence, currently, KGE is being used extensively to tackle problems such as KG completion, entity recognition, and link-based clustering [24, 33, 34].

The core idea underlying KGE is to create a vector for each entity and each relation in the KG, then define a set of score functions that are used to measure the space distance of two entities relative to the type of the relation in the low-dimensional embedding vector space. The aim is to capture latent properties of the semantics in the KG so that like entities and like relations will be represented with similar vectors, and those not semantically connected are detached. The embedding of a KG is learned via training a neural architecture over a KG, and usually includes three main steps [35], namely: (i) encoding entities into dispersed points in the semantic space, and encoding relations as vectors; (ii) scoring function or model-specific function that is used to assemble the information combing from a triple; and (iii) applying an optimization procedure represented by the loss function in which the objective is defined and minimised during the KG embedding process. Table 2 shows the most popular of the many models which are commonly used for KG embeddings.

---

[8] https://spacy.io/
[9] http://corenlp.run/
[10] https://allennlp.org/
[11] https://www.ibm.com/cloud/watson-natural-language-understanding

*Table 2: Examples of well-known KG embedding models*

| Model | Description | Scoring Function |
|---|---|---|
| **TransE** [36] | learns the representation of both the entities and relations as vectors in the same low dimensional semantic space. Hence, for a golden triple $(h,r,t)$, **TransE** treats the relation $r$ as a translation in the embedding space so that $h + r \approx t$, when $(h,r,t)$ holds ($t$ should be the closest to $h + r$), otherwise $h + r$ should be away in distance from $t$. | $f_{TransE} = -\|e_h + e_r - e_t\|_n$ |
| **DistMult** [37] | is an extension and a simplification to RESCAL [38] and is based on the bilinear model. In this model, the relation is encoded as diagonal (single vector) using the trilinear dot product. | $f_{DistMult} = r^T(h \odot t)$ |
| **ComplEx** [39] | this is an extension to DistMult model by introducing complex-valued embeddings, where the scoring function is based on the trilinear Hermitian dot product in $C$. | $f_{ComplEx} = Re(\langle e_r, e_h, \overline{e_t} \rangle)$ |
| **HolE** [40] | a compositional vector space model that learns compositional vector space representations of entities and relations through incorporating the strength of RESCAL as well as the simplicity of **DistMult.** | $f_{HolE} = w_r \cdot (e_h \otimes e_t) = \frac{1}{k} F(w_r) \cdot (\overline{F(e_h)} \odot F(e_t))$ |
| **ConvE** [41] | is a neural link prediction model that uses deep, multi-layer, conventional and fully connected layers of nonlinear features to tackle the interactions between input entities and relations. | $f_{ConvE} = \langle \sigma(vec(g([\overline{e_h}; \overline{e_r}] * \Omega))W))e_t \rangle$ |
| **ConvKB** [42] | incorporates conventional neural networks to represent the concatenation of entities and relations, which increases the learning ability of latent features. | $f_{ConvKB} = concat(g([e_h, e_r, e_t] * \Omega)) \cdot W$ |

## 3.5 Knowledge Graph Evaluation

The proliferation of massive KGs poses a question regarding the quality of the embedded knowledge (i.e., entities and relations), and whether these facts precisely convey the intended real-world concepts interlinked via their relationships. Therefore, ascertaining the completeness and correctness of the constructed KG is crucial to determine its "fitness of purposes" [43] for various downstream applications, and to deal with the uncertainty in the data quality [44].

In a domain-specific KG, the absence of a complete and accurate KG poses a challenge to the evaluation process. This is because collecting all true facts about a certain domain of interest is not a trivial task (if not impossible). Therefore, various attempts, commonly known as KG Augmentation/Completion techniques, have been undertaken to augment the KG with new facts presented by new potential entities and/or new relations. To ensure data quality, these efforts are subject to correctness and completeness evaluation measures. In particular, according to the new and known true facts, the evaluation can be carried out by using classification accuracy and ranking metrics such as Hits@N and Mean Reciprocal Rank (MRR), Accuracy, Precision, Recall, and F-score [5, 45]. These metrics are amongst various other measures that are currently being incorporated to evaluate the KG construction and completion in terms of the factuality of the embedded entities as well as their relations. Table 3 lists some of the well-known evaluation metrics.

Evaluations of KG constructions have also been carried out by means of case studies and domain experts [46, 47].

*Table 3: Examples of well-known KG Evaluation metrics*

| Metric | Description | Formal definition | |
|---|---|---|---|
| **Hits@N** | indicates the number of elements in the ranking vector retrieved from the model is positioned in the top ($N$) locations. | $Hits@N \begin{cases} \sum_{i=1}^{|Q|} 1, if\ rank(s,p,o)_i \leq N \\ 0, \quad otherwise \end{cases}$ | where $rank(s,p,o)_i$ refers to the rank of a positive element $i$ against a list of negative elements, $T$ is a set of test triples and $(s,p,o)$ is a triple $\in T$. |
| **MRR** | is a function that computes the mean of the reciprocal of elements embodied in a vector of rankings. It is used as a measure to evaluate the system performance against the retrieved elements. | $MRR = \frac{1}{|T|} \sum_{i=1}^{|Q|} \frac{1}{rank(s,p,o)_i}$ , | |
| **MR** | refers to the mean rank of the correct test facts/triples embodied in a vector of rankings (i.e., the average of the predicted ranks). | $MR = \frac{1}{|T|} \sum_{i=1}^{|Q|} rank(s,p,o)_i$ | |
| **Accuracy** | specifies the accuracy of the incorporated embedding model in making a correct prediction. Accuracy is the ratio obtained between the accurate predictions (i.e., $TP + TN$) and the overall inferred predictions ($FN + TP + FP + TN$). | $Accuracy = \frac{TP + TN}{FN + TP + FP + TN}$ | True-Positives (TP): refer to the true facts that are classified by the model as true statements. False-Positives (FP): refer to the false facts that are classified incorrectly as true statements. True-Negatives (TN): refer to the false facts that are classified correctly as false statements. False-Negatives (FN): refer to the true facts that are classified incorrectly as false statements. |
| **Precision** | refers to the proportion of those facts that were classified accurately as positive and they are actually positive. | $Precision = \frac{TP}{TP + FP}$ | |
| **Recall** | recall indicates the proportion of true positive facts were correctly classified as positives facts. | $Recall = \frac{TP}{TP + FN}$ | |
| **F- measure** | (a.k.a. f-measure/ f-score) is a harmonic measure used to provide a trade-off between precision and recall | $F - score = \frac{Precision \cdot Recall}{Precision + Recall}$ | |

## 4. Domain-specific KGs

This section reviews various domain-based KGs that were discussed in the literature. These domains are Healthcare, Education, ICT, Science and Engineering, Finance, Society and Politics, and Travel. Appendix A includes seven tables, each with a summary of the literature for each designated domain. These tables demonstrate the specific KG usage, KG construction algorithm(s), the resources used to feed the KG, whether KG embedding techniques were incorporated, the evaluation approach, and the limitations of each designated work.

### 4.1 Healthcare

Recently, the Healthcare sector has gained much attention, particularly with coronavirus 2019 (COVID-19) pandemic continues to rattle the world. Therefore, there is a notable consensus in both industry and academia to consolidate efforts to overcome the challenges of this vital sector [48]. KGs offer the healthcare sector technical means to derive meaningful insights from voluminous and heterogeneous healthcare data [49, 50]. The examined papers in Healthcare domain can be categorised into the following subdomains.

**Generic healthcare** – Health data mining by means of a KG approach was followed in the literature. Gatta et al. [51] presented a library in R that was developed for process mining in the medical domain. The library is designed to encode the extracted processes in the form of directed graphs, which can be then interpreted and visualised by domain experts. Another important effort that integrated plausible reasoning with fine-grained biomedical ontologies to tackle data incompleteness problem in generic health data was undertaken by Mohammadhassanzadeh et al. [47]. The authors proposed Semantics-based Data analytics (SeDan) framework that performs an exploratory and plausible analysis of the KG using plausible OWL extension and query rewriting algorithm. The framework incorporates various knowledge bases including the DrugBank, Disease Ontology, and the large-scale semantic MEDLINE database (SemMedDB). Rastogi et al. [52] framed the personal health KG as a combination of context, personalization, and integration with other knowledge-bases. Their study indicated that the literature on personalised health-related KGs is inadequate and lacks a unified standard representation to depict the designated domain. Incorporating health KGs for Query Answering (QA) system was discussed by Sahu et al. [53]. The authors offered a system that can be used to search for various health-based KGs to obtain a set of healthcare-related response sub-graphs. Incorporating KG in the medical domain to benefit QA applications was also discussed in [54].

**Diseases –** The advances of constructing KGs to conceptualise health-related problems and diseases were discussed in the literature. Rotmensch et al. [55] constructed a KG that captures diseases and symptoms related entities form 273,174 electronic medical records. The authors incorporate Google Health Knowledge Graph (GHKG) and created a KG that embodies diseases and symptoms and relationships between them. Constructing KGs that can describe depression was provided by Huang et al. [56]. In particular, they attempted to generate a sub-graph, that describes depression disorder, obtained by parsing a variety of large knowledge sources such as PubMed, Medical Guidelines, DrugBank, Unified Medical Language System (UMLS) etc. Evaluating the robustness of a constructed KG in healthcare is of utmost significance to ensure the quality of the inferred knowledge in this sensitive domain. In this context, Chen et al. [57] presented a methodology to measure and evaluate the robustness of knowledge in terms of diseases and symptoms captured from existing health knowledge graphs as well as records of patient visits to the Beth Israel Deaconess Medical Center (BIDMC). Addressing the temporal dimension in KG creation is an important dimension in healthcare. Ma et al. [58] established a temporal KG that can be used for cognitive episodic memory. This temporal KG was initially derived from the Integrated Conflict Early Warning System (ICEWS) dataset as well as Global Database of Events, Language and Tone (GDELT). Their work was different from other seminal works by generalizing four significant static KGs embedding to 4-dimensional temporal/episodic KGs. Also, two novel generalizations of RESCAL were proposed and discussed. Application of KGs in healthcare and medical domains was demonstrated in other relevant tasks such as fraud, waste, and abuse Detection [59], drugs similarity [18], drug repurposing [60], clinical decision support systems [61], and medical recommender systems [62, 63].

**Healthcare management –** Constructing a KG to benefit health management and to address current health-related problems and chronic diseases were proposed in the literature [64-67]. For example, Huang et al. [64] suggested a KG construction model that benefits people seeking knowledge regarding a healthy diet. The authors proposed a domain ontology as an underlying structure of a diet KG. The KG was then enriched with entities extracted from a set of healthcare websites using Conditional Random Fields (CRF), Support Vector Machine (SVM) and Decision Tree (DT) algorithms. Another effort was carried out by Haussmann et al. [65] who proposed an integrated KG (FoodKG) that embodies knowledge on healthy food, recipes, and nutrition. The authors ensured the credibility of the obtained knowledge by adopting RDF Nanopublication specification [68]. On the same research direction, an inclusive healthy diet KG was also constructed by Chi et al. [66]. The reported KG integrated five key concepts that included food material, dish, nutritional element, symptom, and the crowd. Through semi-automatically extraction approach, the proposed model was capable to collect and import entities captured from a set of online resources using various NLP and machine learning algorithms. Modeling food domain-specific KGs were also implemented in [69-71]. Further, tackling challenges in healthcare systems leveraging KGs technologies was discussed in [72-74].

4.2 Education

The construction and usage of educational KGs have been extended recently due to the significance of KGs application to the learning systems as well as the abundance of pedagogical data [75]. The following subdomains are inferred from the collected education-related papers.

**Teaching and classroom resources –** KGs have proven ability to foster learning [76] and been used in popular massive open online course (MOOC) platforms [77, 78]. For example, Chen et al. [79] presented K12EduKG, a KG constructed based on K-12 educational subjects. Domain-specific educational data (*Chinese curriculum standards of mathematics*) was the source of knowledge that was used in K12EduKG. Concepts and relations are identified and imported into K12EduKG using CRF model and probabilistic association rule mining. Su and Zhang [80] designed a KG schema that can accommodate educational Big data. Their KG was enriched by using two large datasets, namely subject teaching resources and an online encyclopedia resource. Another attempt has been undertaken to build a KG (MathGraph) that can be used to solve high school mathematical exercises including mathematical derivation and calculation [81]. The authors of [81] used MathGraph, that was initially constructed with the help of crowdsource, to embody dissimilar mathematical objects, operations and constraints. KnowEdu [77] is one of the important efforts in educational KG design and construction. By using standard curriculum and learning assessment data as data sources, and by using neural network models for concepts and relations identification, KnowEdu is created to facilitate learner's cognitive and educational process.

**Education management –** Aliyu et al. [82] presented an approach for implementing a KG to be used for course allocation scheduling and management. Although the evaluation of the constructed KG was inadequate, the work is promising toward this important research direction. Liu et al. [83] and Lian et al. [84] provided approaches that adopted graph-based relational learning for concept prerequisite learning in education domain. The former introduced an automatic technique for prerequisites prediction by inferring directed graph at the course and concept levels. The latter incorporated active learning to the concept prerequisite learning problem. Modelling internal control in higher education using KG technology was discussed in [85]. The author proposed a KG that can conceptualise the internal control policy in the higher educational institutions. Despite the author's attempt to demonstrate the utility in a visualisation task, the overall mechanism to construct and evaluate the designated KG is inadequate and inferior. Applying KGs to benefit education management was also outlined in [86].

**Educational technologies –** Designing a KG that can be used to depict academic networks was discussed in [87]. The authors proposed a model of scientific publication management that can integrate scientific metadata in terms of academic entities. The model embodies a KG which conveys the relationship between research entities and research topics. Incorporating KG embedding techniques in the process of constructing the educational KGs is inadequate, particularly those leverage the rich literals of the designated KGs, thus Yao et al. [88] attempted to tackle this issue by reporting on a model for embedding learning of educational KGs. With the use of three experimental KGs in the education domain, the authors demonstrated the significance of the proposed model when processing educational KGs. In particular, authors of [88] presented a method that can jointly learn embeddings built on pre-trained structural (i.e. TransE) and literal embedding vectors (i.e. BERT). Evaluating the utility of the educational KG by means of visualisation analysis was rarely discussed in the literature. An attempt in this direction was undertaken by Sun et al. [89] who integrated an education KG and demonstrated its utility by carrying out visual analysis so as to provide a better understanding to its topological structure.

## 4.3 ICT

KGs have been widely used to improve several subdomains related to Information and Communication Technology as follows:

**Cybersecurity** – Detecting and preventing cyberattack is inevitable to ensure providing continuous and uninterrupted services. Interestingly, various cybersecurity-related KGs have been introduced and developed. For example, [90] presented a practical approach for cybersecurity. They first developed a domain ontology that put forward a technique to construct the cybersecurity KG. Then they proposed a quintuple model that was used to obtain new knowledge using the path-ranking algorithm. Deng et al. [91] discussed another cybersecurity-related KG that was constructed to serve students who seek concepts in this domain. Despite the Adhoc mechanism to construct their KG and the absence of a benchmark comparison for utility evaluation, the line of research is important per se; providing KGs that facilitate personalized learning and benefit education is highly recommended [92]. Kiesling et al. [93] followed a bottom-up approach to build their cybersecurity KG using National U.S. Vulnerability Database (NVD) and set of security online references. The authors demonstrated the effectiveness of the developed KG by means of two case studies in vulnerability assessment and intrusion detection systems. Cybersecurity KGs were also reported in [94-96].

**Software development** – Software development is a sophisticated process that encompasses an array of challenges and decisions to be made in a timely manner [97]. Hence, the software engineering domain has also benefited from the propagational use of domain-specific KGs due to their efficacy to store and manage relevant entities and relations of high complexity. For example, Nayak et al. [98] developed a KG that was used to extract test cases that would assist in the functional requirements gathering process. As a backbone schema, the authors designed an ontology for software testing and applied a series of NLP tools including Constituency Parse Tree (CPT) to mine and populate the KG with test cases. Schindler et al. [99] introduced a KG (SoftwareKG) that embodies information pertaining to the software mentioned in academic articles of social science. SotwareKG depicts various aspects of the software including its availability, source and links with other knowledge repositories. Designing a framework to industrial software design and development processes was proposed by [100]. The authors applied a knowledge-driven QA system for parameters searching and can be also used for carrying out variable calculation and ontology reasoning. The proposed model integrated the constructed KG with an SQL database and efficiency is demonstrated in certain industrial scenarios. Fu et al. [101] made use of IT crowdsourcing services to construct a KG (ITServiceKG) to improve existing IT services IR system. The authors implemented a learning-to-rank model (Gradient Decision Tree) that was leveraged to re-rank the obtained results, thereby attaining much relative search results. Constructing KGs to improve software engineering practices and internal processes were also addressed in [102-105].

**Telecommunication** – Defining and conceptualizing the structure of telecommunication networks can be explicitly delivered through incorporating KGs. Aumayr et al. [106] benefited from the unique structure of KG to build a graph that can be used to solving issues that confront telecom operators. The authors populated the KG with entities captured form community knowledge in forms of telecom and products documentations, online sites, engineering reports, etc. The aim was to build an automated system to improve network incident management processes and to provide better customer service. Krinkin et al. [107] proposed a telecommunication network monitoring model by means of incorporating a domain-specific KG at the top of the telecommunications service domain ontology. In particular, the authors encompass an array of components that are integrated into monitoring cable television operator networks. These components are: the billing model, user access rights, network topology and application hierarchy, and the cable television operator network service model.

**Internet of Things (IoT) –** In IoT integrating heterogeneous access of electronic devices poses a momentous challenge. Hence, the underlying structure of the KG offers a promising solution to bridge the gap between IoT devices. For example, Xie et al. [108] proposed an IoT KG that was used in a new layer to map IoT devices, thereby unifying the communications of all devices. Many studies further elaborated on KGs and their advantages to the IoT ecosystem [109-111].

4.4 Sciences and Engineering

Applying semantic web technologies and ontologies in natural sciences has proven successful leveraging the formal knowledge representation and the semantic web languages that can model rich and complex knowledge of natural sciences [112-116]. Examples of subdomains that benefited from the KG technology are reviewed as follows:

**Chemistry –** The utility of semantic analytics has been validated in providing a formal representation to chemical data, thereby increasing the sharing and interoperability of such data [117]. Incorporating KG has extended these endeavours and provided a platform where information can be integrated from multiple chemical kinetic systems, and offered an automatic method to comprehend chemical mechanisms to perform complex chemical-related semantic queries [118]. For example, [119] proposed an integrated system that used KG to demonstrate interoperability in cross-domain applications that compass combustion as well as to address the problem of data inconsistencies in chemical reaction mechanisms. Krdzavac et al. [120] designed a domain ontology (OntoCompChem) as an underlying structure of a KG that was used to demonstrate quantum chemistry calculations.

**Biology –** The continuous propagation of the throughput data in molecular biology has necessitated the introduction of formal representation of this knowledge domain. For example, Humayun et al. [121] developed a mechanistic KG to depict the heme's interactome, an important factor of diverse biological processes. Choi et al. [122] carried out benchmark comparison amongst a selection of KG embedding models in a relational discovery task. Prior to the implementation of the incorporated KG embeddings, the authors constructed a domain-specific KG that was populated with entities and relations captured from heterogeneous public data sources. These bio-related data sources are PubMed database[12], Comparative Toxicogenomics database (CTD)[13], The Biological General Repository For Interaction datasets(BioGRID)[14], and the human disease database: (MalaCards)[15].

**Geology –** Incorporating KGs on geological data has proven effectiveness and enhanced the interconnectivity between such data sets. Zhu et al. [123] showcased the use of KG in an intelligent system for deep mining of geological data. The authors constructed the KG using Baike.com and local geological documents. Constructing a KG from geoscience documents was discussed in [124]. The authors made use of an integrated corpus composed of a geology dictionary and the *Terminologies and Classification Codes of Geology and Mineral*

---

[12] http://www.ncbi.nlm.nih.gov/PubMed/

[13] http://ctdbase.org/

[14] https://thebiogrid.org/

[15] https://www.malacards.org/

*Resources (*TCCGMR) [125] to enrich the KG incorporating CRF-base geological word segmentation model. The utilisation of KG to conceptualise geosciences were further detailed in [126-128].

**Engineering –** More advanced technical solutions have been also importing KGs to build sophisticated systems in various fields of engineering [129]. For example, Myklebust et al. [130] demonstrated the implication of using a domain-specific KG and KG embeddings to improve the ecotoxicological effect prediction in the Norwegian Institute for Water Research (NIVA). In particular, the authors designed TERA KG to integrate information captured from dissimilar resources relevant to ecotoxicology and risk assessment domain (e.g. ECOTOXicology database (ECOTOX)[16] and NCBI taxonomy[17]). The construction of TERA KG was carried out by using LogMap ontology alignment system [131] to index and align the ECOTOX and NCBI vocabularies. Yan et al. [132] designed KnowIME (KG's Intelligent Manufacturing Equipment), a knowledge-based integration system for manufacturing equipment such as lathes, conveyors and robots. A domain-specific KG was constructed and augmented using CRF method from heterogeneous equipment-related data. Industrial adoption to KGs to enhance customer satisfaction and user experience was undertaken by [133]. The authors designed two KGs for evolutionary Smart Product– Service System (Smart PSS) development. The constructed KGs resulted from data obtained from open-source knowledge, prototype specifications, and user-generated textual data. The KGs were then used to bringing expert knowledge, thereby solving issues related to cost-effectiveness that exist in the current knowledge supply for Smart PSS.

Electric power artificial intelligence systems have also contributed to the KGs construction and augmentation. For example, Fan et al. [134] proposed an approach to construct the dispatch KG for the power grid to semantically describing behavior of dispatchers. They follow semi-automated labelling to construct a power corpus, then BiLSTM-CRF model was used to extract entities and indicate the dispatching behavior relationship patterns. Yang et al. [135] also leveraged the KG schema to collect and integrate data from various power assets. They aimed to deliver a unified multi-source heterogeneous knowledge base that contains power transmission and transformation assets. Engineering is a wide spectrum of domains that have found KGs advantageous in several important applications. This is evident in various other engineering fields such as, nuclear engineering [136], marine engineering [137], Photonics engineering [138], Nanotechnology Engineering [139], Ceramics engineering [140], and Geomatics engineering [141].

4.5 Finance

The finance sector is a pillar of any successful business. It is the driver for businesses to take opportunities and make revenue. Accordingly, researchers commonly draw great attention to this domain by discovering new venues for continuous improvements. Intuitively, KGs, as being a powerful tool for various applications, have been constructed to enrich several subdomains of Finance [142]. The following discussion sheds the light on recent works pertaining to two important subdomains of Finance, namely financial investments and fraud detection.

**Financial investments –** Liu et al. [143] leverage a domain-specific KG to carry out stock market forecasting on the renowned companies. Their work also comprised a deep learning approach and proven effectiveness when integrated with the constructed KG on the prediction task. Another attempt was commenced by Fu et al. [144]. The authors introduced a stochastic optimisation algorithm, genetic programming, and generalised crowding which are all integrated into a model for market return prediction using financial KG. Cheng et al. [145] proposed

---

[16] https://cfpub.epa.gov/ecotox/
[17] https://www.ncbi.nlm.nih.gov/taxonomy

KG-based event embedding framework that is designed for event-driven quantitative investment. In particular, the constructed KG (named FinKG) and the implemented embeddings performs learn informative representations based on both the relations of event argument and the lead-lag relations amongst the entire KG. Liu et al. [146] demonstrated the use of a KG embedding framework to predict stock prices using news sentiment analysis. Although the authors did not provide much discussion on the validity of the mechanism followed to construct the KG, the utility was demonstrated in the prediction task. In the same line of research, Long et al. [147] integrated trading data, public market information and investor's records to construct a KG that is incorporated to model the market and its features. The KG was then embedded using node2vector approach and used in a deep neural network model for forecasting trends in stock prices. Authors of [148] depicted the relevance of using a knowledge-empowered model on event representation and stock prediction. Zhang et al. [149] supported the aforementioned endeavours by proposing an approach to detect short-term stock price movement. The authors developed an enterprise KG and designed a top-up power vector model and influence propagation model. The aim was to compute the effect of a specific relationship from the relevant enterprise. The construction of the KG involved incorporating Named Entities Recognition (NER) and Neural Relation Extraction (NRE) for entities extraction and Convolutional Neural Network (CNN) for relation inference. Stock management and stock prediction tasks have been also discussed in [150-154].

**Fraud Detection –** Financial fraud detection is an important area of financial risk management. KGs have been leveraged to establish approaches that can be used to stop such criminal activity. For example, Wang et al. [155] used a finance KG as a basis for label propagation algorithm to detect online fraud. Their model embodies a partition algorithm that is used to distinguish fraudulent groups of users. They argued that fraudulent users tend to position a close distance, whereas normal users commonly exist in isolated tense or firmly connected groups. Zhan et al. [156] also proposed a model that was applied in the fraud detection domain. In particular, the authors designed a call network KG that is enriched with call historical data and loan transactional data. Although the mechanism followed to construct the KG and the evaluation metrics are inadequate and inferior, the research topic is important and emphasises the significance of combining KGs with machine learning techniques to detect fraud in finance [157, 158].

4.6 Society and Politics

Societies were initially formed men who have intuitively established political systems by making formal and informal decisions concerning the production, distribution, and the use of various resources [159]. Thus, entities in terms of people, organisations, and resources have been involved to establish various forms of relations which exist among them [160]. The formal representation of KGs provides an excellent mean to conceptualise relationships in social sciences and politics.

**Social Science –** The current advances in information and communication technologies have made a qualitative leap and have created new venues where people can exchange thought, ideas and interest [97, 161-163]. This ICT revolution is embodied into various electronic means depicted by the emergence of social media. The data created by such platforms are propagating posing questions on the quality of the data being generated by these platforms [164-167]. Hence, there is a vital need to study these platforms and provide ground truth of trustworthy data sets [166, 168-170]. Tchechmedjiev et al. [171] introduced ClaimsKG, a knowledge graph of fact-checked claims originated from International Fact-Checking Network (IFCN). The purpose of ClaimsKG is to enable users searching for true facts of a certain entity. Similarly, it can be used to infer false facts of people, organisations, etc. Nguyen et al. [172] created a KG that embodied social events decomposed from social media using Independent Component Analysis (ICA) and the SocioScope Knowledge Graph (SKG) model. ICA is used to cluster social events obtained from a matrix of collected hashtags. This was followed by using the SKG model do automatically construct event-driven KGs from Twitter data. Huang et al. [173] reported a KG that can be used in social media to detect entity morphs (aliases that are commonly used to conceal the identity of a certain entity). The developed KG includes the real entity linked with all identified morphs mansions. A topic modelling algorithm( i.e. CorrLDA2), as well as SVM models, were used in the KG construction process. Various comparative studies were undertaken to demonstrate the effectiveness of the proposed approach.

**Politics –** Processing social data to infer domain knowledge that can be used to design a political KG was also discussed in the literature. For example, our previous work [174] developed a credibility-based Politics and applied various KG embedding techniques to validate the KG's utility. In particular, BBC politics ontology was incorporated and extended as a backbone schema for the Politics KG. Then, various domain knowledge inference tools were used to enrich the KG with political entities captured from the collected datasets. Finally, several KG embedding models were implemented and tested over a set of link prediction, clustering, and visualisation tasks. The work proposed by [175] crossed with the former model as both efforts emphasized the significance of adding trust aspect in the process of designing and constructed a KG. In particular, the authors presented POLARE, an ontology that conceptualised the political system, and built a KG based on the provided ontology schema so as to be used for a better understanding the existing relations between agents in the political system in Brazil. Another attempt to detect and infer Political ideology was proposed by [176] whereby the authors introduced an opinion-aware KG that was used for conducting political ideology forecasting. The model integrated knowledge captured from social media, DBpedia and ideological books corpus. Rudnik et al. [177] made use of Wikidata to semantically annotating news articles. The annotated articles were then fed into a predefined event-oriented KG that was used for semantic-based search engine. News recommendation by means of a KG was also examined in [178]. The authors applied a filtering method to eliminate irrelevant relations from the currently incorporated and propagated KG. Microsoft Satori KG was used as a backbone knowledge-based and enriched with entities captured from MSN news corpus. Also, the authors introduced article topic entities and the collaborative edges as two new categories of information to be embedded in the original graph. Mehdi et al. [179] designed a KG embedding approach based on socio-scholarly KG which embodies scientific artifacts on social good. The developed system incorporated various KG embedding approaches so as to retrieve, for a given entity (i.e., publication, author, domain and venue), all related and semantically-matched entities. Other studies were further established systems over KGs to benefit news and journalism domain [180-182].

**Culture –** Modelling culture and history of societies have been also conceptualised using domain-specific KGs. For example, Liu et al. [183] developed a KG that depicts ancient Chinese history and culture. The author constructed the KG employing Baidu Encyclopedia[18] as the knowledge source, BiLSTM-CRF for entity recognition, and DeepKE (developed by Zhejiang University) for relation inference. Constructing cultural knowledge bases that benefit from domain ontology and cultural KGs was elaborated further in [184-186].

4.7 Travel

Travel is one of the key domains which availed from the growing use of KGs. In particular, Tourism and Trafic/Transportation are amongst the subdomains that have largely employed KGs.

**Tourism –** KGs have been constructed and used in touristic QA systems [187] or in tourism recommender systems to recommend personalized attractions [188] and the best accommodation prices [189]. For example, Kärle et al. [189] gathered data about Tirol region at Austria and constructed "Tirol Tourism" KG to benefit applications such as eCommerce. The constructed KG was fed by entities and relations extracted from Destination Management Organizations (DMOs) and Geographical Information Systems (GIS) and other tourism and accommodation-related websites. Establishing a touristic KG for China was introduced in [190]. The authors designed the domain-specific KG by using data obtained from Chinese encyclopedia KG and unstructured web pages. For entity alignment, the authors made use of Skip-Gram Model to fitch relative entities during the knowledge acquisition phase. Another attempt to construct a KG for Chinese tourism was undertaken by [187]. The aim was to build a QA system based on the implemented KG of tourism. The authors followed a proposed entity recognition algorithm for entity extraction and utilised CNN model for relation inference. In a different context, Calleja et al. [191] created "DBtravel" KG over the Spanish entries of Wikitravel[19]. The authors followed GATE[20] pipeline that encompasses three internal processes, namely: (1) tokenizer, (2) sentence splitter and (3) named entity recognition. Employing KGs in the tourism domain has been highly active recently and used in various applications [192-194].

---

[18] https://baike.baidu.com/

**Transportation and traffic –** Applying KGs in transportation and traffic has also obtained much attention recently due to the population growth, air pollution, and other sophisticated embedded issues that require intelligent systems to resolve them efficiently [195]. For example, Zhou et al. [196] introduced a model to predict urban traffic congestion. In this model urban KG was constructed from miscellanies static and dynamic raw urban data. The authors applied CNN to model spatio-temporal correlation between each indicated region. One of the well-known KGs that was applied to the US transportation system is ATMGRAPH [197] which was built at the top of the NASA ATM Ontology (ATMONTO) [198]. M. Keller [197] in ATMGRAPH combined an array of structured aviation data obtained from the large part by US federal agencies. The developed KG conceptualises the US National Airspace System by incorporating entities describing, airspace infrastructure, flights, and flight operating conditions. Detecting traffic events by employing KG was proposed in [199]. The authors built a KG named ITSKG (Imagery-based Traffic Sensing Knowledge Graph) that was used to comprehend traffic patterns based on stationary traffic camera imagery data. Further, Wang et al. [200] followed a semi-automated KG construction approach based on China railway electrical accidents data. The purpose of the work is to analyse and identify the faulty equipment of railway electrical accidents in China as well to infer patterns and trends from such data. In a different context, Zhang et al. [201] have expanded the exertions in the maritime transportation by applying KG to build a knowledge representation of the regulations depicted by International Maritime Dangerous Goods Code (IMDG Code). They aimed to facilitate the access and retrieval of detailed guidelines embedded in IMDG. Knowledge-based models in transportation domain were also benchmarked [202] and others were used in travel-related intelligent systems to facilitate information sharing and interoperability [203, 204].

4.8 Summary
The aim of this study was to review the latest publications concerned with KG construction approaches in seven domains of knowledge, namely: healthcare, education, ICT, science and engineering, finance, society and politics, and travel. A close examination of these interesting domains reveals important areas of research that can greatly benefit from KG technology. This study not only demonstrates the popularity of incorporating KGs to arrive at solutions for real-life problems, but also illustrates how KGs have proven to be an effective overall solution to mitigate complexity, ensure flexibility, and establish a common-ground topology whereby data can be integrated from different sources. This is clearly articulated in the selected articles in which there is a consensus that KG technology facilitates the integration of information captured from various sources. For example, we observe several efforts where KGs have been constructed using heterogenous resources to build consolidated QA and recommender systems in healthcare [47, 56, 65], to discover relations in biodata [122], to predict ecotoxicological effects in environment engineering [130], to detect fraud in finance [156], and to discover and visualise political relationships [174, 176]. These attempts demonstrate how knowledge can be obtained from different sources, which perhaps exist in different formats, and can be then fed into one coherent schema to be formally used to conceptualise the designated domain. At the same time, we observe that several other works, particularly in the sciences and engineering domain, have not taken advantage of this important feature of KGs, but have employed limited data sources to feed their constructed KGs. Further observations and findings from this study are discussed in the next section.

---

[19] https://wikitravel.org/
[20] https://gate.ac.uk/

## 5. Findings from the Survey

The review of KG construction approaches which are drawn from academic works in seven domains reveals a correlated array of limitations and deficiencies related to the following summarised points:

A) **KG data quality, privacy, and credibility**: The reviewed articles have not consistently used standardized and appropriate data quality measures, particularly with the construction of large-scale KGs. Various approaches imported data collected from noisy and low-quality data sources (such as social media) with little consideration given to the credibility or privacy of generated information. This is also seen in electronic medical records from which it is difficult to collect data due to privacy and confidentiality constraints. This raises a question regarding the quality and robustness of KGs that are constructed from such data repositories. Despite some attempt to tackle data quality issues [43, 46], the endeavours in this direction have been inadequate. Blockchain technology presents a promising breakthrough in this regard as it provides continuous verification and advanced audit trails of all transactions. Blockchain technology does not only provide a complete mechanism to ensure data quality, integrity, and accuracy; it also has a secure and tamper-proof architecture for data storage, which would be extremely beneficial for healthcare information. Future research should be steered toward designing methodologies that establish best practices to construct *immutable* domain-specific KGs leveraging both KG and blockchain technologies.

B) **Knowledge resources and semantic expansion**: Semantic Web technologies and Linked Open Data (LOD) have opened the door wide to improving several domain applications [205]. KGs are extensions to these efforts and are commonly associated with LOD projects as they enrich the semantics of data by providing conceptual representations of concepts and entities [10]. Therefore, the interoperability of information is facilitated by relevant entity interlinking obtained from other KG repositories, thereby constructing multi-modal KGs. However, the approaches examined in the designated domains have demonstrated limitations in achieving interoperability of information. In particular, semantic expansion/broadening techniques were insufficiently incorporated to benefit from the openly available vocabularies and curated semantic repositories. Essentially, the underlying structure of KGs is designed to pave the way for data integration, unification, and information sharing and usability. Research in this direction should be reinforced to ensure that the essence of KGs adheres to FAIR (Findable, Accessible, Interoperable, Reusable) principles[21].

---

[21] https://www.go-fair.org/fair-principles/

C) **KG construction algorithms**: Deriving meaningful knowledge from a diversity of data formats is not a trivial task; it involves extracting facts about entities and their potential relationships, which requires a correlated array of various Information Extraction (IE) techniques and sophisticated Natural Language Processing (NLP) approaches. The review of the selected articles revealed only limited discussions on techniques used for entity recognition and/or relation extraction. Most of these studies either neglected to specify the algorithm(s) used in the KG construction including techniques for entity and relation extraction (e.g. [135, 144, 155]) or presented few details and poor rationale for using such techniques (e.g. [51, 90, 107, 145]). From the analysis, we also identified an existing research gap - there were no consolidated methodologies for automating the process of constructing a domain-specific KG that facilitates the selection of suitable algorithms for the designated techniques (involving both NLP methods and data-driven approaches [206]). Such methodologies would not only improve KG construction practices; they could also make the knowledge broadly available for access by both humans and machines.

D) **Time-aware KGs**: The dynamic nature of knowledge is highly correlated to contextual situations; hence, various facts that describe entities might change over time. Therefore, the temporal dimension should be integrated into KGs. Various currently propagated KGs are static but temporary, and do not consider the time factor [207]. This applies to both open-world and domain-specific KGs. Neglecting the dynamic nature of knowledge harms the quality and correctness of facts contained in the KGs and might lead to poor decision making that is based merely on such data sources. Consequently, despite some exceptions such as Wikidata and YAGO in which certain facts are already endowed with the time information, the construction of KGs should consider the validity period of facts [208].

E) **KG evaluation**: During the process of constructing KGs, incorrect facts in terms of entities and/or relations could be captured. This process is prone to errors, particularly in regard to the information derived from mix-quality data sources. As mentioned previously, it is vital to consider the credibility of the data source, as is the evaluation of the overall construction process. In fact, among the reviewed articles, the evaluation of KGs was the predominant weakness. For example, some studies carried out a superficial and subjective evaluation of the KG construction with no incorporation of concrete evaluation metrics [134, 144, 199]. Another thread of efforts attempted to involve theoretically-proven evaluation metrics to systematically measure KG completion and KG correctness approaches. The former approaches were commonly measured using recall, precision, and f-measure [64, 66, 80, 90]. The latter incorporated accuracy and/or area under the ROC curve (AUC) [7, 57, 147]. Nevertheless, our study endorses the work conducted by [7] in that, providing holistic techniques which simultaneously improve the quality of KGs in dissimilar domains, is still an open research problem. Further, the absence of testbeds and benchmark datasets has prevented the community from undertaking an appropriate and fair evaluation of techniques used for KG creation [209]. Lastly, in their evaluation methodologies, researchers should consider various other data quality indicators including completeness, believability, relevance, objectivity, consistency, understandability, etc. [210].

F) **Computing performance in Big KGs**: There is a notable consensus amongst the research community that the conventional technologies used to process and analyse the continuous propagation large-scale datasets are no longer adequate [211]. Machine Learning and Artificial Intelligence have become the preferred methods for the processing and analysis of Big data, by means of which the hoped-for added value is obtained. Large-scale KGs (i.e., those with trillions of triples) adhere to various Big data features. Volume is not the only feature of Big KGs; they can be described by the variety in data sources, the velocity of data generation, veracity of data quality, volatility of data currency and availability, etc. Despite some growing efforts to integrate large-scale KGs in Big Data processing [212], there is a need for a deeper and congruent integration of Big Data technology infrastructures and sophisticated statistical models required for reasoning over domain-specific KGs, and this remains an open research venue.

G) **Domain-specific KG Reasoning**: KG reasoning aims to provide new interpretations and conclusions from constructed KGs. It is a mechanism whereby new facts can be inferred from an existing KG [213]. KG embedding-based approaches (including tensor decomposition, distance, and semantic matching models) have gained considerable attention in the literature due to their scalability and efficiency in accommodating large-scale KGs so as to deduce a generalizable context about the KG that can be applied to infer new relations [8, 214]. This study discloses a lack of incorporating KG embedding techniques into the examined approaches. Applying KG embeddings extends these efforts and tackles the deficiencies that might occur in the KG construction process, which subsequently leads to incomplete graphs. To address this issue and to establish a technical ground for conducting relational reasoning in AI systems, KG embeddings should be integrated and implemented in domain-specific KG construction models.

H) **Availability of domain-specific KGs**: Unlike open-world and generic KGs, most of the constructed and populated domain-specific KGs are not published on the Web and thus are not accessible to other researchers. The paucity of such annotated and enriched data silos makes the process of benchmark comparison more difficult, as well as limiting reusability, interlinking and interoperability. At the same time, there is a scarcity of relevant experts who can perform data annotation. For example, in the healthcare domain, there is a lack of detailed clinical entities and relations. This is due to the shortage of both clinical repositories [50] and clinicians to conduct data annotation [215]. This issue has also been highlighted by researchers in Cybersecurity [216], Finance [217], and Social Science [218], to name a few.

I) **Room for further research on domain-specific KGs**: this study attempts to capture a snapshot of recent advances in KGs in certain domains of knowledge. Due to limitations of space, not all KGs construction attempts in every domain and its subdomains could be considered. Hence, there are still various specific areas of the selected high-level domains that can exploit KG technology. For example, there is a need for technical solutions to respond to COVID-19 as a critical catalyser of various technological applications. KG technology can provide a shared conceptualised structure which can be used to map all dispersed facts stored in various data silos which are relevant to this pandemic. Integrating information from various available resources pertaining to COVID-19 can help to bridge the knowledge gap and provide a better understanding of this disease. COVID-19 has also given impetus to the adoption of new technologies for remote and online education. The propagation of online learning resources has established a space where domain-specific KGs can be employed to alleviate the problem of selecting relevant resources to benefit both teachers and learners. KG technology can be applied to solve an array of problems in any domain where the data is collected from different sources. In these applications, KG technology is an excellent means of achieving interoperability, unification, interlinking, and data integration.

# 6. Conclusion

The widespread use and the prevalence of the Internet have led to the rapid increase in data volume which has necessitated the development of advanced data analytics that are capable of handling the propagation and the heterogeneity of such data. Knowledge graphs continue to dominate as a distinctive form of data representation and knowledge inference, and are core activity of several industrial applications. The great amount of interest in this technology is due to its underlying structure that is built on a formal conceptual representation that is depicted by a domain ontology. Therefore, domain-specific KGs have been constructed and used to tackle several real-life problems in dissimilar domains. Yet, to date, there has been no consensual definition of a domain-specific KG. Moreover, the current methods used to construct and evaluate domain knowledge graphs are far from perfect.

This paper is the first to present an inclusive definition of domain-specific KG. Further, it provides an in-depth analysis of the current state-of-the-art knowledge graph construction research in seven domains of knowledge. The efforts undertaken in each research domain are discussed, and the shortcomings and limitations of these efforts are considered, along with future research opportunities which we hope will motivate researchers in this specific area.

# Appendix A. Tables

*Table A.1: Overview of KG approaches in Healthcare domain*

| Ref. | Sub-domain | KG Usage | Construction Algorithm(s) | KG Resource(s) | Embedding Technique(s) | Evaluation Measure(s) | Limitation(s) |
|---|---|---|---|---|---|---|---|
| [51] | Generic | Medical process mining | Markov Models and Careflow Mining, | Event logs | N/A | Case study | <ul><li>Inadequate evaluation,</li><li>inadequate discussion on the KG construction approach</li></ul> |
| [47] | Generic | QA | Plausible reasoning | BioASQ, DrugBank, Disease Ontology, and SemMedDB | N/A | Domain expert's verification | <ul><li>Insufficient evaluation,</li><li>evaluating the performance of query rewriting algorithm does not exist</li></ul> |
| [55] | Diseases | Identification of diseases and symptoms across the collected medical records | LR, NB and Bayesian network modeling | Custom, GHKG | N/A | Precision and Recall | <ul><li>Inadequate to infer correct causal relations,</li><li>concept extraction requires further elaboration</li></ul> |
| [56] | Diseases | QA - Depression disorder | NLP tools | PubMed, DrugBank, DrugBook, and UMLS | N/A | Use cases | <ul><li>Lack of proper evaluation,</li><li>insufficient use of other important medical repositories</li></ul> |
| [57] | Diseases | Benchmark comparison-Diseases and symptoms | LR, NB, and (noisy OR) | BIDMC, GHKG | N/A | AUC, F1 score, and AUPRC | <ul><li>Limited data sources,</li><li>poor causal inference methods</li></ul> |
| [58] | Diseases | Episodic memory - Inductive learning | Manual integration | GDELT and ICEWS | Tucker, RESCAL, HolE, ComplEx, DistMult, ConT and Tree | MRR, and AUPRC | <ul><li>Poor generalization due to the timestamps not observed in the training dataset,</li><li>KG embeddings are prone to overfitting (too many parameters)</li></ul> |
| [64] | Healthcare management | Healthy diet recommendation | CRF, SVM and DT | Healthcare websites | N/A | Precision, Recall, and F1-score | <ul><li>Limited to food and dietary,</li><li>Chinese language only,</li><li>inadequacy to prove utility in link prediction and other knowledge discovery tasks</li></ul> |
| [65] | Healthcare management | Food recommendation, and QA systems | Lexical similarity and string matching | DBpedia, USDA, Recipe1M and FoodOn KG | word2vec, and FastText | Case study and F1-score | <ul><li>Lack of evaluation on both KG constructions and incorporated embeddings techniques,</li><li>inadequate meaningful representations for food recommendation</li></ul> |
| [66] | Healthcare management | Food recommendation | NLP tools, CRF, SVM, NB, LSTM, and KNN | China Food Composition and Online health websites, | N/A | Precision, recall, F1-measure, and Questionnaire | <ul><li>Limited data sources,</li><li>lack of proper recommendations,</li><li>directed solely for the Chinese context</li></ul> |

*Table A.2: Overview of KG approaches in Education domain*

| Ref. | Sub-domain | KG Usage | Construction Algorithm(s) | KG Resource(s) | Embedding Technique(s) | Evaluation Measure(s) | Limitation(s) |
|---|---|---|---|---|---|---|---|
| [79] | Teaching and classroom resources | K-12 Education | CRF and probabilistic association rule mining | Chinese curriculum standards of mathematics | N/A | AUC | • Insufficient evaluation, <br> • narrow scope of the proposed KG and its applications |
| [80] | Teaching and classroom resources | Learning assessment and recommendation | Bootstrapping construction strategy and BERT-BiLSTM-CRF | Subject teaching resources, Baidu Encyclopedia, and DBPedia | N/A | Precision, Recall and F1 measure | • Limited demonstration on the utility of the constructed KG including assessment of student learning, etc. |
| [81] | Teaching and classroom resources | Solving high school mathematical exercises | Complex, Triangle, Conic and Solid | Crowdsourcing and domain experts | N/A | Accuracy, Precision, Recall and F1 measure | • Limited resources used for KG construction, <br> • limited targeted audience |
| [77] | Teaching and classroom resources | Learning assessment | RNN and probabilistic association rule mining | Pedagogical data and learning assessment data | N/A | AUC, Precision, Recall and F1 measure | • Lack of elucidating the effects of KG in other settings, <br> • their schema is relatively hard to be provided for conducting benchmark comparison |
| [82] | Education management | QA and course allocation scheduling | Adhoc | Structural educational information system | N/A | Case study | • Poor evaluation measures, <br> • limited KG resources, <br> • narrow scope |
| [85] | Education management | Internal policy control conceptualization and visualization in higher education | Adhoc | CNKI database | pTransE | mean Silhouette | • Limited data sources, <br> • limited application scope, <br> • poor evaluation metrics, <br> • KG embedding was not properly demonstrated and evaluated |
| [88] | Educational Technologies | Link Prediction | Adhoc | Knowledge Forest, Wikipedia | TransE and BERT | Mean Rank and Hits@10 | • Insufficient structural and literal embedding models were used |
| [87] | Educational Technologies | Decision-making in academia | NLP tools, SVM, NB, and LR | Web of Science, Engineering Village, and EBSCO | N/A | F-score | • The limited scope of KG (can be expanded to include instructors and their metadata), <br> • no domain ontology is provided as a base for the proposed KG, <br> • limited evaluation measures |

*Table A.3: Overview of KG approaches in ICT domain*

| Ref. | Sub-domain | KG Usage | Construction Algorithm(s) | KG Resource(s) | Embedding Technique(s) | Evaluation Measure(s) | Limitation(s) |
|---|---|---|---|---|---|---|---|
| [90] | Cybersecurity | Cyberattack detection | Stanford NER | Enterprise data and security websites | N/A | Precision, Recall, and F1 | • Narrow KG construction approaches,<br>• limited evaluation with current state-of-the-art KGs in the designated domain |
| [91] | Cybersecurity | QA and RS for education | Adhoc and NLP tools | Wikipedia | N/A | Case study and survey | • Limited data sources,<br>• no benchmark comparison,<br>• no backbone ontology schema |
| [93] | Cybersecurity | Vulnerability Assessment and Intrusion Detection | Adhoc and RML Rules[22] | NVD and security online sites | N/A | Case studies | • No benchmark comparison,<br>• inadequate rationale on the construction approach |
| [98] | Software development | Test cases extraction | CPT and CRF | Software documents, requirement statements and test reports | FastText algorithm | Accuracy, Precision, Recall, F1 | • Transfer learning can replace the incorporated NER model, thereby using pre-trained datasets instead,<br>• Inadequate validation to the collected requirement statements and past test reports |
| [99] | Software development | Investigating Software Usage in the Social Sciences | bi-LSTM and bi-LSTM-CRF | DBpedia and PLoS | N/A | Case study | • Narrow to PLoS which affected the target domain,<br>• evaluation is inadequate as no benchmark comparison was undertaken,<br>• the automatic linking process with DBpedia requires a further scrutiny |
| [101] | Software development | Info. Retrieval (IT crowdsourcing services) | Adhoc | StackOverflow, Wikipedia and crowdsourcing platforms | N/A | MRR, P@K, Recall | • Ranking modeling system can be enhanced with incorporating neural networks |
| [100] | Software development | Industrial software design and development processes | Adhoc | Generic(public databases and unstructured data sources) | N/A | Case study | • Limited domain-based data sources,<br>• unable to provide recommendation in complex and domain-specific situations,<br>• poor retrieval performance |
| [106] | Telecom | Telecom incidents managements | NLP tools | Network incident documents | N/A | Case study on information reduction and discovery | • No discussion on the effectiveness of the incorporated algorithms/tools for KG construction,<br>• no benchmark comparison with similar state-of-the-art |
| [107] | Telcom | Television operator network monitoring | Adhoc | Data obtained by monitoring systems | N/A | Case study | • Limited discussion on KG construction and propagation,<br>• no rationale on using the designated ontology schema,<br>• inadequate evaluation method |
| [108] | IoT | Bridging gaps between e-devices in IoT | Adhoc | oneM2M[23] | N/A | Case study | • Inadequate discussion on concept and relation extraction approaches,<br>• limited in data sources and application scope |

---

[22] https://github.com/carml/carml
[23] https://www.onem2m.org/component/rsfiles

*Table A.4: Overview of KG approaches in Science and Engineering domain*

| Ref. | Sub-domain | KG Usage | Construction Algorithm(s) | KG Resource(s) | Embedding Technique(s) | Evaluation Measure(s) | Limitation(s) |
|---|---|---|---|---|---|---|---|
| [119] | Chemistry | Combustion chemistry modelling | J-Park Simulator | Linked open data | N/A | Query and simulation systems | • Limited data sources,<br>• lack of human and machine–interaction tools |
| [122] | Biology | Biodata relational discovery | Manually | PubMed, CTD, BioGRID, and MalaCards | TransE, PTransE, TransR and TransH | Hits@10 | • Poor KG construction approach,<br>• unsatisfactory evaluation of the incorporated KG embedding models |
| [123] | Geology | Geological IR system | HanLP24 and association rule analysis | Baike.com | N/A | Case study | • Limited data sources.,<br>• undefined underlying structure,<br>• inadequate evaluation to the utility of KG. |
| [124] | Geology | Chinese geology Knowledgebase | CRF | Geology dictionary and TCCGMR | N/A | Case study | • Limited data sources,<br>• inadequate concept and relation extraction,<br>• no benchmark comparison with similar state-of-the-art |
| [132] | Manufacturing engineering | Intelligent manufacturing equipment recommendation | CRF | Equipment-related data (e.g. Baidu Encyclopedia, etc.) | N/A | Case study (Information richness and effectiveness) | • Limited to level of stand-alone equipment,<br>• scattered manufacturing data that harden data acquisition,<br>• poor accuracy |
| [133] | Design engineering | Evolutionary Smart Product–Service System Development | Domain experts and NLP toolkits | Smart PSS prototype, user generated textual data, misc. medical websites | N/A | Showcase | • Proposed solutions for the personalized requirements are someway generic and oversimplified,<br>• unoptimized incorporated algorithms that led to poor complexity in terms of time and space |
| [130] | Environment engineering | Ecotoxicological effect prediction | LogMap | ECOTOX, NCBI and Wikidata | TransE, DistMult, and HolE | Accuracy, Precision, Recall, and F-score | • Limited data sources; can be enhanced with information about species and compounds |
| [134] | Electrical engineering | Improving power dispatching process. | BiLSTM-CRF | Power dispatching texts | N/A | Subjective evaluation | • Small, static and non-diversified dataset,<br>• poor evaluation mechanism |
| [135] | Electrical engineering | Improving utilization of power assets information | Adhoc | PMS and ERP | N/A | Case study | • Limited evaluation measures,<br>• limited data sources,<br>• limited KG structure and application |
| [219] | Oil and gas | Intelligent search engine for oil and gas | BiLSTM-CRF | Center of Oil and Gas | N/A | Precision and Recall | • Limited data sources,<br>• lack of optimization to the embedded search engine. |

---

*Table A.5: Overview of KG approaches in Finance domain*

| Ref. | Sub-domain | KG Usage | Construction Algorithm(s) | KG Resource(s) | Embedding Technique(s) | Evaluation Measure(s) | Limitation(s) |
|------|-----------|----------|---------------------------|----------------|------------------------|----------------------|---------------|
| [143] | Investment | Stock market prediction | Adhoc | Tomson Reuters, CNN | word2vec, TransE, Neural network | Accuracy and F1-score | • Limited resources led to insignificant scale of the trained dataset,<br>• vague KG construction approach (no algorithm is indicated),<br>• reporting the utility of the KG embedding was inadequate |
| [144] | Investment | Market return prediction | Adhoc | Shanghai Stock Exchange and WIND Financial Terminal | N/A | Subjective evaluation based on 29 component stocks | • KG construction in terms of entities and relations was not properly addressed,<br>• missing KG schema,<br>• poor evaluation to the resultant KG |
| [145] | Investment | Event-driven quantitative investments | OpenIE v5.1 | Financial news websites | N/A | Micro – F1, Weighed – F1 | • Poor discussion on KG construction mechanism/algorithm,<br>• lack of justification on the use of customized evaluation metric |
| [146] | Investment | Stock price volatility prediction | Rule-based named entity recognition | Financial news websites and UQER | TransR | Accuracy (prediction task) | • Large scale KG that can affect the performance of TransR embedding model,<br>• the KG construction validity was not scrutinized prior using it in the prediction task |
| [147] | Investment | Stock price trend prediction | Adhoc | Chinese securities companies | node2vec | AUC | • KG modelling was neither illustrated nor validated,<br>• sentiment analysis can be integrated to obtain better performance results |
| [149] | Investment | Stock price movement direction prediction | NER, NRE, and CNN | Online financial news | N/A | Acc, MCC, and FM | • Lack of temporal dimension,<br>• limited data sources,<br>• discussion on KG construction is inadequate |
| [155] | Financial risk management | Fraud Detection | N/A | Orange Finance | N/A | AUC, Precision, Recall and F1 measure | • Unindicated KG construction mechanism/algorithm,<br>• inadequate demonstration to the KG schema |
| [156] | Financial risk management | Fraud Detection | Adhoc | Loan transaction data and call history | word2vec | Precision Acceleration-Recall curve | • Challenging data acquisition and might lead to poor KG construction and prediction performance accordingly,<br>• poor evaluation to the implemented KG |

*Table A.6: Overview of KG approaches in the Society and Politics domain*

| Ref. | Sub-domain | KG Usage | Construction Algorithm(s) | KG Resource(s) | Embedding Technique(s) | Evaluation Measure(s) | Limitation(s) |
|---|---|---|---|---|---|---|---|
| [171] | Social science | QA (fact-checked claims) | TagMe tool | International Fact-Checking Network websites | N/A | Case study | • Limited data sources, <br>• poor KG construction algorithm, <br>• limited evaluation metrics |
| [172] | Social science | Detecting and tracing social events. | SKG model | Social media (Twitter) | N/A | Precision, Recall, and F-measure | • Complexity can be improved by increasing periods and applying ICA on other cases, <br>• limited data sources, <br>• schema of the KG is inadequate to properly representing the relationships |
| [173] | Social science | Identify Entity Morphs | CorrLDA2 and SVM | DBpedia, Yago, and Freebase | N/A | Precision, recall, and F-measure | • Too many useless generate morphs can be avoided by adopting certain heuristic algorithms, <br>• morphs can be also extended to cover not only people but events and other entities |
| [179] | Social science | Social good recommender system | Adhoc | Academic research papers | TransE, ComplEx, TransH, TransR, TransD, DistMult and RESCAL | Mean rank and hits@k | • Limited conceptual presentation of the KG in terms of entities and relations, <br>• results can be improved by using reinforcement learning |
| [174] | Politics | Link prediction, clustering, and visualisation | Adhoc using IBM Watson NLU | BBC Politics ontology, Wordnet, Google KG and light-wight ontologies. | TransE, DistMult, ComplEx, HolE, ConvE, and ConvKB, | Hits@N, MMR, Accuracy, Precision, Recall and F-score | • Limited discussion on the utility of the KG in a practical real-life example, <br>• limited discussion on the KG construction algorithms. |
| [175] | Politics | Trustworthy claims fact in the political system | Adhoc/ nanopublication model25 | Se Liga na Politica(SLNP) | N/A | Case Study | • Limited direct relations between political organizations, <br>• lack of fine-grained patterns in the political agent domain, <br>• the utility of the incorporated provenance dimension was not properly validated |
| [176] | Politics | Political ideology detection | Holistic lexicon-based approach | DBpedia, Twitter and ideological books corpus | N/A | Accuracy | • Imperfect use of evaluation strategy and metrics, <br>• no KG Embedding was undertaken, thus, no rationale provided on the benchmark comparison, <br>• limited data sources |
| [177] | Politics | News articles retrieval | SpaCy | Wikidata | N/A | N/A | • Limited data sources, <br>• evaluation metrics were not provided |
| [178] | Politics | Politics news recommendation | Adhoc | MSN News corpus and Microsoft Satori | TransE | AUC, NDCG @10, and F1-Score | • Sophisticated model construction that hardens the process of usability and interoperability |
| [183] | Culture | QA and RS for Chinese ancient history and culture. | BiLSTM-CNN-CRF26 and DeepKE | Baidu Encyclopedia | N/A | Precision, recall, and F1-score | • Inadequate use of named entity extraction techniques, <br>• limited data sources |

---

[25] http://nanopub.org/wordpress/
[26] BiLSTM-CNN-CRF: Short-Term Memory-Convolutional Neural Networks-Conditions Random Field

*Table A.7: Overview of KG approaches in Travel domain*

| Ref. | Sub-domain | KG Usage | Construction Algorithm(s) | KG Resource(s) | Embedding Technique(s) | Evaluation Measure(s) | Limitation(s) |
|---|---|---|---|---|---|---|---|
| [189] | Tourism | Touristic IR system for Tirol at Austria | Adhoc wrapper software | DMOs, GISs, Feratel[27], Infomax[28], web maps[29], Outdooractive[30] and waldhart[31] | N/A | Case Study | • Scalability; hard to provide a wrapper for each source when it maps to Schema.org, <br>• data collection is subject to error-prone, <br>• poor KG construction evaluation |
| [190] | Tourism | Chinese tourism-domain knowledge service | NLP -Skip-Gram Model | Sogou-T[32], Chinese Wikipedia dump[33], and Zhishi.me[34] | N/A | Accuracy | • Inadequate evaluation measures, <br>• lack of benchmark comparisons with other related KGs |
| [187] | Tourism | KG-QA in tourism | Adhoc entity recognition algorithm and CNN | Manual collection and NLPCC2016KBQA dataset[35] | N/A | Accuracy | • Inadequate discussion on rationale on entity and relation extraction, <br>• limited evaluation metrics |
| [191] | Tourism | Spanish tourism-oriented knowledge service | Adhoc based on GATE pipeline | Wikitravel | N/A | Precision, recall and F-measure | • Construction of the KG led to import noisy data, <br>• limited data sources. |
| [196] | Transportation and traffic | Urban traffic congestion | Adhoc | Beijing traffic data and meteorological data. | N/A | Accuracy, and F1 | • Limited discussion on KG construction including entity and relation extraction approaches |
| [199] | Transportation and traffic | Traffic image feature extraction | Adhoc | Imagery data | N/A | Subjective evaluation | • Poor evaluation approach, <br>• traffic cameras are prone to errors (weather, lighting, maintenance, etc.) |
| [200] | Transportation and traffic | Railway Electrical Accident Analysis | Adhoc and BiLSTM-CRF | China Railway electrical accidents data | N/A | Precision, Recall and F1 | • Limited data sources, <br>• inadequate discussion on utility of the constructed KG, <br>• lack of benchmark comparison |
| [201] | Transportation and traffic | Knowledge integration of maritime dangerous goods. | Adhoc | IMDG Code | N/A | Case study | • Poor discussion on entity and relation extraction algorithms, <br>• lack of proper evaluation metrics to validate the utility of the proposed KG |

---

[27] https://www.feratel.com/

[28] https://www.infomax.de/

[29] https://general-solutions.eu

[30] https://www.outdooractive.com

[31] https://www.waldhart.at/

[32] https://www.sogou.com/labs/resource/t.php

[33] https://dumps.wikimedia.org/zhwiki/

[34] http://openkg.cn/dataset/zhishi-me-dump

[35] https://github.com/huangxiangzhou/NLPCC2016KBQA